\documentstyle[12pt,aasms]{article}

\begin{document}

\title{SEQUENTIAL STAR FORMATION IN TAURUS MOLECULAR CLOUD 1}
\author{Tomoyuki Hanawa\altaffilmark{1} and
Satoshi Yamamoto\altaffilmark{2}}
\affil{Department of Astrophysics, School of Science, Nagoya University,
Chikusa-ku, Nagoya 464-01, Japan}

\author{Yasuhiro Hirahara\altaffilmark{3}}
\affil{Department of Pure and Applied Sciences, College of Arts and Sciences,
The University of Tokyo, Meguro-ku, Tokyo 153, Japan}

\altaffiltext{1}{I: hanawa@a.phys.nagoya-u.ac.jp}
\altaffiltext{2}{Present address: Department of Physics,
The University of Tokyo, Bunkyo-ku, Tokyo 113, Japan}
\altaffiltext{3}{Present address: Department of Earth and
Planetary Sciences, School of Science, Nagoya University, Chikusa-ku,
Nagoya 464-01, Japan}

\begin{abstract}
     We discuss the fragmentation of a filamentary cloud
on the basis of a 1-dimensional hydrodynamical simulation of a self-gravitating
gas cloud.  The simulation
shows that dense cores are produced with a semi-regular interval
in space and time from one edge to the other.  At the
initial stage the gas near one of the edges is
attracted inwards by gravity and the accumulation of the gas
makes a dense core near the edge.  When the dense core
grows in mass up to a certain amount, it gathers gas from the other
direction.  Accordingly the dense core becomes isolated from the
main cloud and the parent
filamentary cloud has a new edge.  This cycle repeats and the
fragmentation process propagates towards the other edge.
The propagation speed is a few tens of percent larger
than the sound speed.   According to the theory, the age difference
for the northwest-most and southeast-most cores in TMC-1 is estimated to be
0.68~pc/0.6~km~s$^{-1}$ = $ 10 ^6 $~y.
The estimated age difference is consistent with that
obtained from the chemical chronology.
\end{abstract}


\section{INTRODUCTION}

\bigskip

      The star formation process seems to propagate by several mechanisms
on both large and small scales (see, e.g., a review by Elmegreen 1992).
A classical example of propagating
star formation is the linear sequences of stellar subgroups
in Orion OB associations (\cite{bla64}).  A larger scale example
is the super giant shell in the Large Magellanic Cloud (\cite{wm66}).
The shell consists of HI clouds and bright blue stars, and surrounds
two supernova remnants and stellar associations.  Similar shells
are also seen in our Galaxy.  In these examples, massive stars
trigger star formation in the neighborhood through supernova
explosion and UV radiation.

      The star formation process may also propagate in a dark cloud.
Recently Hirahara et al. (1992) have found that dense cores are
successively formed.  According to them,
a filamentary cloud, Taurus Molecular Cloud 1 (TMC-1)
is elongated in the NW-SE direction and contains
five dense cores called A-E from NW to SE.
The cores have different chemical ages and are older in NW than in SE.
The age difference is of the order of $ 10 ^6 $ y.
This indicates propagation of dense core formation.

     If cores A - E are formed successively, it is likely that
a causal mechanism works there.  However it could be
neither supernova explosion nor UV radiation from young stars;
no influential stars are found in the vicinity of TMC-1.
Although various models are proposed for sequential star formation
on large and small scales, most of them invoke young massive stars.
They cannot be applied to TMC-1.

     In this paper we propose a new mechanism for sequential core (star)
formation which operates in a filamentary molecular cloud.  In our model,
formation of a dense core triggers that of another core.
Once a core is formed, the gravitational field is changed and
accordingly the gas flows.  Another core is produced by the gas
flow and the cycle repeats.  Using a 1-dimensional (1D)
hydrodynamical model we demonstrate
how this mechanism works.

     In section 2 we construct a 1D
model for the dynamics of a filamentary
cloud.  Some characteristics of the 1D model are investigated
analytically.  It is shown that the 1D model can explain the fragmentation
of a filamentary cloud due to the self-gravity of the gas.
Numerical simulations of the 1D model are shown in section 3.
They demonstrate that cores are produced sequentially with semi-regular
temporal and spatial intervals in a filamentary cloud.
The apparent propagation speed of fragmentation is somewhat larger than
the effective sound speed.
In section 4 we summarize observations so far reported for TMC-1 and
apply our model to TMC-1.
The apparent
propagation speed of fragmentation in TMC-1 is of the order of the turbulent
velocity.  Since the filament length and velocity dispersion are
0.68 pc and 0.6 km~s$^{-1}$ in TMC-1, the time scale for the propagation
of fragmentation is estimated to be $ 10 ^6 $ yr.
This estimate is consistent with the age of cores calculated from
their chemical abundances.
In section 5 we discuss the dynamical and chemical evolution of
a filamentary cloud during fragmentation.

\section{1-DIMENSIONAL MODEL}

\subsection{Model Equations}

     When a filamentary cloud fragments, the gas motion is
mainly in the direction parallel to the filament axis.  Thus,
we approximate the gas motion to be 1-dimensional.
The equation of motion and the equation of continuity are
expressed as
\begin{equation}
\rho \Bigl( { \partial v \over \partial t } \, + \,
v { \partial v \over \partial z } \Bigr) \, + \,
{\partial P \over \partial z } \, + \, \rho {\partial \psi \over
\partial z} \, = \, 0 \; ,
\end{equation}
and
\begin{equation}
{\partial \rho \over \partial t} \, + \, {\partial \over \partial z}
( \rho v ) \, = \, 0 \; ,
\end{equation}
respectively, where $ v $, $ \rho $, $ P $, and $ \psi $ are
the velocity along the filament axis, the density, the pressure, and
the gravitational potential, respectively.  The Poisson equation is
arranged to be
\begin{equation}
{ \partial ^2 \psi \over \partial z ^2 } \, - \, { \psi \over R ^2 }
\, = \, 4 \pi G \rho \; ,
\end{equation}
where $ G $ and $ R $ are the gravitational constant and a length
scale for the filament radius.  The derivative in the
radial direction in the Poisson equation is replaced with a constant
coefficient $ \lbrack $ the second term in the  left hand side of
equation (3) $ \rbrack $.  Equations (1) thorough (3) denote the
gas motion in a filamentary cloud under the 1-dimensional approximation.

\subsection{The Filamentary Cloud Model and Its Stability}

      In this subsection we discuss a model for a infinitely long
$ ( \partial / \partial z \, = \, 0 ) $ filamentary cloud and its
stability.  A steady state solution of equations (1), (2), and (3) is
expressed as
\begin{equation}
\rho (z) \, = \, \rho _0 \, = \, {\rm const.} \;
\end{equation}
\begin{equation}
 v (z) \, = \, 0 \; ,
\end{equation}
and
\begin{equation}
\psi (z) \, = \, - 4 \pi G \rho _0 R ^2 \; .
\end{equation}
We consider a small perturbation around this equilibrium.
The density perturbation is assumed to be in the form of
\begin{equation}
\rho \, ( \, z , \, t ) \; = \, \rho _0 \, + \,
\rho _1 \, {\rm exp} \, ( i k z \, + \, i \omega t ) \; .
\end{equation}
Then the dispersion relation is expressed as
\begin{equation}
 \omega ^2 \, = \, - \, 4 \pi G \rho _0 \,
{ k ^2 R ^2 \over 1 \, + \, k ^2 R ^2 } \, + \, c _s {} ^2 k ^2 \; .
\end{equation}
Equation (8) is qualitatively similar to the approximate dispersion
relation for a filamentary  cloud (\cite{han93}),
\begin{equation}
 \omega ^2 \, = \, - \, 4 \pi G \rho _c \,
{ k  H \over 1 \, + \, k H } \, + \, c _s {} ^2 k ^2 \; ,
\end{equation}
where the equilibrium density distribution is assumed to be
\begin{equation}
 \rho _0 \, = \, \rho _c \, (1 \, + \, r ^2 / 8 H ^2 ) ^{-2} \; .
\end{equation}
This similarity justifies our 1D approximation and suggests
$ R \, \simeq \, H $.

     In our 1D approximation, the model is unstable whenever
\begin{equation}
 4 \pi G \rho _0 R ^2 \; > \; c _s {} ^2 \; .
\end{equation}
We think that inequality (11) is satisfied in a gravitationally bound
filamentary  cloud.  For a 2-dimensional model of a filamentary
cloud there is a condition of magnetohydrodynamical
equilibrium in the radial direction;
\begin{equation}
 4 \pi G \rho _c H ^2 = c _s {} ^2 + 2 \Omega _c {} ^2 H ^2
 +  { B _c {} ^2 \over 16 \pi \rho _c } (1 \, + \, \cos ^2 \theta )
\; ,
\end{equation}
where $ \Omega _c $, $ B _c $ and $ \theta $ denote the angular velocity
at the filament center, the magnetic field density at the center,
and the pitch angle of the magnetic fields at $ r \, = \, \infty $
(see Hanawa et al. 1993 for further details).  Equation (12) means
that the gravitational force should be equated with the sum of the thermal
and magnetic pressures and the centrifugal force.  If a filamentary cloud
is permeated with magnetic fields and/or rotates around the axis,
inequality (11) is fulfilled.  For example,
TMC-1 has velocity gradient along the
minor axis which can be ascribed to rotation around the major axis
(e.g., \cite{oww88}).
In the following we assume inequality
(11) except when otherwise stated.

      The growth rate and wave number of the most unstable mode are
derived to be
\begin{equation}
 ( - i \omega ) _{max} \, = \, \sqrt{ 4 \pi G \rho _0} \, - \, c _s / R
\, ,
\end{equation}
and
\begin{equation}
  k _{max} \, = \, {1 \over R} \, \sqrt{
{ R \sqrt{ 4 \pi G \rho _0 } \over c _s } \, - \, 1 } \; ,
\end{equation}
from equation (8).  The values of $ ( -i \omega ) _{max} $,
$ k _{max} $, and $ \lambda \, \equiv \, 2 \pi / k _{max} $ are
summarized in Table 1.  Equations (13) and (14) are used to interpret
the 1D nonlinear simulation of fragmentation.

\subsection{Periodically Distributed Cores}

      Equations (1) - (3) have a more general steady state solution where
the density changes periodically.  In hydrostatic equilibrium
the density is expressed as
\begin{equation}
 \rho (z) \, = \, \rho ( 0 ) \, {\rm exp} \, \Bigl\{ -
 \, { \psi (z) \, - \, \psi (0) \over c _s {} ^2 } \Bigr\} \; .
\end{equation}
Substituting equation (15) into equation (3) we obtain
\begin{equation}
{d ^2 \psi \over d z ^2} \, - \,
{\psi \over R ^2} \, = \, 4 \pi G \rho (0) \,
{\rm exp} \, \Bigl\{ - \, {\psi (z) \, - \, \psi (0) \over c _s {} ^2 }
\Bigr\} \; .
\end{equation}
In the following we assume that the density has the minimum value,
$ \rho (0) \, = \, \rho _{\rm min} $ at $ z \, = \, 0 $, and
hence the gravitational potential is the highest at $ z \, = \, 0 $, with
$ \psi (0) \, = \, \psi _{\rm max} $.
Integrating equation (16) multiplied by $ d \psi /d z $ we obtain
\begin{eqnarray}
\Big\vert {d \psi \over d z} (z)  \Big\vert ^2 & = &
\Big\vert { \psi (z) \over R } \Big\vert ^2
-  \Big\vert { \psi (0) \over R } \Big\vert ^2 \nonumber \\
 - 8 \pi G & \rho ( 0 ) & c _s {}^2 \, \Bigl\{
{\rm exp} \Bigl( - { \psi (z) \, - \, \psi (0) \over c _s {} ^2 }
\Bigr) \, - \, 1 \Bigr\} \; .
\end{eqnarray}
Since $ d \psi / d z $ is expressed as a function of $ \psi $,
we can obtain $ z $ as a function of $ \psi $ for given $ \rho (0) $
and $ \psi (0) $ which satisfy
\begin{equation}
 \, - 4 \pi G \rho (0) R ^2 \, \le \, \psi (0) \, < \, 0 \; .
\end{equation}
When $ \psi (0) \, = \, - \, 4 \pi G \rho (0) R ^2 $, the density and
gravitational potentials are spatially constant.

\section{NUMERICAL SIMULATION}

\subsection{Numerical Methods}

      The differential equations (1) - (3) are integrated numerically in
Lagrangian coordinates with a fully implicit scheme.  We introduce
a Lagrangian coordinate, $ s $, defined by
\begin{equation}
 ds \, = \, \rho ( z ) \, d z \; .
\end{equation}
Using $ s $ we rewrite equations (1) - (3) into
\begin{eqnarray}
 \Bigl( { d z \over d t } \Bigr) _s \, & = & \, v \; , \\
\Bigl( { d v \over d t } \Bigr) _s \; & = & \;
- \, {1 \over \rho _c} \Bigl( { \partial P \over \partial s } \Bigr)
\; - \; \ g \; , \\
P \; & = & c _s {} ^2 \rho \; , \\
\rho \, & = & \, \rho _c \, \Bigl( {d z \over d s } \Bigr) ^{-1}
\; , \\
{\partial g \over \partial s} \; & = & \; {\rho _c \psi \over \rho R ^2} \;
+ \; 4 \pi G \rho _c \; , \\
{\partial \psi \over \partial s} \; & = & \; ( \rho _c / \rho ) g \; .
\end{eqnarray}
We followed the time evolution of a Lagrangian mesh point,
$ z _i \, (0 \, \le \, i \, \le \, n)$ using the difference equations,
\begin{equation}
 z _j = z _j ^{\rm (old)} + \Delta t \, v _j \; ,
\end{equation}
\begin{equation}
 v _j = v _j ^{\rm (old)} + \Delta t
\Bigl( - c _s {} ^2 \, { \rho _{j+1/2} -  \rho _{j-1/2}
\over \rho _c \, \Delta s }
- g _j \Bigr) \, ,
\end{equation}
where
\begin{eqnarray}
\rho _ {j+1/2} \; & = & \;
\rho _c \, \Bigl( { z _{j+1} \, - \, z _j \over \Delta s }
\Bigr) ^{-1} \, \\ \nonumber
\psi _{j+1/2} \; & = & \;
{ \rho _{j+1/2} R ^2 \over \rho _c } \,
{ g _{j+1} \, - \, g _j \over \Delta s } \\
& & - \; 4 \pi G \rho _{j+1/2} R ^2 \; ,
\end{eqnarray}
\begin{equation}
g _j \; = \; { \rho _{j+1/2} \, + \, \rho _{j-1/2} \over 2 \rho _c } \,
{ \psi _{j+1/2} \, - \, \psi _{j-1/2} \over \Delta s } \; .
\end{equation}
We take the fixed boundary condition,
\begin{equation}
 z _0 \, = \, {\rm const.} \; \; {\rm and} \; \;
z _n \, = \, {\rm const.} \; ,
\end{equation}
at the both ends.
The number of the Lagrangian mesh points is $ n \, = \, 1.1 \times 10 ^4 $
in most of numerical simulations shown in this paper.  The time step
is taken to be $ \Delta t \, = \, 5.0 \times 10 ^{-3} $.

      In the following we take units of $ R \, = \, 1 $,
$ \rho _0 \, = \, 1 $, and
$ 2 \pi G \rho _0 \, = \, 1 $ except when otherwise stated.

\subsection{Standard Model (Model 1)}

      As a typical example we describe model 1 where the sound speed is
taken to be $ c _s \, = \, 1.0 $.  The initial gas distribution is
assumed to be
\begin{eqnarray}
\rho \, & = & \, \, 0.1 \, \rho _0
\hskip 1.0 true cm  ({\rm for} \; -50 \, \le \, z \, \le \, 0) \\
\nonumber \\
& = & \, 1.0 \, \rho _0 \hskip 1.0 true cm
({\rm for} \; 0 \, < \, z \, \le \, 50) \, \\ \nonumber
\end{eqnarray}
and
\begin{equation}
 v \, = \, 0 \; .
\end{equation}
The time evolution of model 1 is shown in figure 1 with grey scale
representation.  The abscissa and ordinate are the $ z $-coordinate
and time, respectively.

     At an early stage with $ t \, \simeq \, 2 $,
the gas near the left edge of the cloud is attracted by the cloud gravity
and flows to right.  A dense core is produced near $ z \, \sim
\, 4 $ at $ t \, \simeq \, 4 $.  The maximum density at $ t \, = \, 6 $ is
$ \rho \, = \, 3.574 $ at $ z \, = \, 5.433 $.
The core is in a quasi-static equilibrium after
$ t \, \ge \, 8 $.  As the core grows in mass, it attracts gas also from
its left side.  The first core becomes isolated from the main cloud at
$ t \, \simeq \, 10 $ and the cloud has a new edge.

     The second core is produced in a similar way at $ t \, \simeq \, 12 $.
Cores are produced sequentially with a time interval of
$ \Delta t \, \simeq \, 8 $ and spatial interval of
$ \Delta z \, \simeq \, 9 $.  The spatial interval is nearly equal to the
wavelength of the fastest growing mode in this model
$ (\lambda _{max} \, = \, 9.763) $.  The apparent
propagation speed of fragmentation, $ v _{frag} \, \equiv \,
\Delta z / \Delta t \, \simeq \, 1.1 $, is a little faster than the
sound speed.  Nakamura, Hanawa \& Nakano (1993) estimated the propagation
speed to be
\begin{equation}
 v _{frag} \, = \, \sqrt{ 2 \omega (k _{max})
 {\partial ^2 \omega \over \partial k ^2} _{k = k _{max}}} \; ,
\end{equation}
from a linear stability analysis (see their Appendix 4).
Substituting equations (13) and (14) into equation (35) we obtain
\begin{equation}
 v _{frag} \, = \, \Bigl( 10 \, - \,
{ 6 R \sqrt{ 4 \pi G \rho } \over c _s } \Bigr) ^{1/2} \, c _s \; .
\end{equation}
Equation (36) gives a good estimate,
$ v _{frag} \, = \, 1.23 \, c _s $ for model 1.

      Figure 2 shows the density and velocity distribution at
$ t \, =  \, 20 $.  The density profile has peaks at $ z \, = \,
9.00 $, 15.73, and 23.30, where the peak density is 18.06, 9.19,
and 2.60, respectively.  The first and second cores are in a
quasi-static state and the velocity is almost constant in each core.
These cores accrete gas from both sides and standing shock waves
are formed at the core surfaces.  The formation of the third core is
in progress.  The first and second
cores attracts each other and merge into a single core at a later stage.
This means that the periodic distribution of gas discussed in subsection 2.3
is also unstable against merging.

      The exact epoch of merging depends on the details of the initial
density distribution and even on the number of the mesh points.
Although a trial computation with fewer mesh points is quite similar
to model 1 in the early stage, the first and second cores merge
earlier in the trial computation.  This is because fewer mesh points give
a larger amplitude of numerical noise.  Compare model 1 with models 4 and 5
for the dependence on the initial density distribution.

\subsection{Dependence on $ c _s $}

      In order to study the dependence on $ c _s $ we constructed
models 2 and 3 where the sound speed is taken to be
$ c _s \, = \, 1.1 $ and 1.2, respectively.  The initial density
distribution of models 2 and 3 is the same as that of
model 1.

       The time evolution of models 2 and 3 are shown in figures
2 and 3, respectively.  Models 2 and 3 are qualitatively similar
to model 1.  The spatial interval between cores is
$ \Delta z \, \simeq \, 10 $ and $ \simeq \, 12 $ for models 2 and 3,
respectively.  The propagation
speed of fragmentation is $ v _{frag} \, \simeq 1.2 $ and 1.4
for models 2 and 3, respectively.  When $ c _s $ is larger,
the cores are wider in separation and fragmentation propagates faster.
These relations qualitatively agree with equations (14) and (36).

\subsection{Dependence on Initial Density Distribution}

      We constructed models 4 through 7 to study the dependence on
the initial density distribution.
The initial density profile has a sharp edge at $ z \, = \, 0 $ in
models 1, 2, and 3.  The sharp edge is introduced for making a
model which is as simple as possible
and unlikely to be realized in interstellar
space. In models 4 through 7  the initial density
increases from 0.1 to 1.0 with a constant density gradient $(d \rho / d z)$
in the region $ - a \, \le \, z \, \le \, a $ where $ a $ is taken
to be $ a \, = \, 2.5 $, 5, 20, and 50 in models 4, 5, 6, and 7, respectively.
The sound speed is taken to be
$ c _s \, = \, 1.0 $ in models 4 through 7.
Figures 5, 6, 7, and 8 show the time evolution of models 4, 5, 6, and 7,
respectively.  Sequential fragmentation also takes place in models
4 through 7.

     Model 4 is not much different from model 1.
A major difference between them is the distance between the
first and second cores at $ t \, = \, 24 $.  It is shorter in model
1 than in model 4.  A small difference in an early stage is amplified
by merging instability.

     Model 5 is also similar to models 1 and 4 except that
formation of the first core is delayed by $ \Delta t \,
\simeq 2 $ and shifted to the right by $ \Delta z \, \simeq \, 3 $ in
model 5.  The delay is due to the low density gradient at the initial
stage and the shift is because the effective edge is shifted to the right
in model 5.  Model 5 is qualitatively similar also in the
formation of the second and third cores to models 1 and 4.
A sharp edge at the initial stage is not essential
for the propagation of fragmentation.

     Model 6 shows also the propagation of fragmentaion.  In model 6
cores are formed both to the left and right of the first core.
Propagation to the left is slower than that to the right.  This is
because the mean density is lower and the self-gravity is
weaker in the left of of the first core.
Propagation to the right terminates at $ t \, \simeq \, 28 $ when
it reaches to $ z \, = \, 50 $, the boundary of computation.
Propagation to the left will terminate after it reaches to a less
dense region (see also figure 6).

     In model 7 the initial density increases from 0.1 to 1
in the region $ -50 \, \le \, z \, \le \, 50 $.  The first core
is formed at $ z \, = \, 50 $ where the initial is the maximum.
Since the reflection boundary is assumed at $ z \, = \, 50 $, framentation
propagates both to the left and right from $ z \, = \, 50 $
although only propagation to the left is shown in figure 8.

     As seen models 4 through 7 propagation of fragmentation is
ubiquitous in the gravitational instability of a long filamentary cloud.
Once a core is formed, subsequet cores are formed in general on the left
and right sides.  When the first core is formed near the edge of a cloud,
propagation to the diffuse side diminishes (see figure 6).
The first core may often be formed near an edge of a filamentary
cloud since the large density gradient thereof triggers
formation of the first core.  In such a case cores are formed from
an edge to the other in a long filamentary cloud.

\subsection{Comparison with 2D Simulation}

     Bastien (1983) followed the non-linear fragmentation of a
non-rotating non-magnetized cylindrical cloud with a 2D hydrodynamical
code for the first time.  The numerical simulation was extended by
Rouleau \& Bastien (1990), Bastien et al. (1992), and Arcoragi et al.
(1992).  They followed the collapse and fragmentation of an initially
uniform cylindrical cloud having  density, $ \rho _0 $,
diameter, $ D $, and length $ L $.

     The qualitative features of
their simulation depend mainly on $ J _0 $, the ratio of the gravitational
to the thermal energies.
When $ J _0 \, > \, J _c $, the cloud is not massive enough to be
gravitationally bound.  When $ J _0 \, > \, J _{\rm spindle} $, the
cloud collapses as a whole into a thin needle and does not fragment.
The case of $ J _c \, < \, J _0 \, < \, J _{\rm spindle} $ corresponds
to our 1D simulation which implicitly assumes hydrostatic balance
in the radial direction.  When $ J _c \, < \, J _0 \, < \, J _{\rm spindle} $,
two condensations are formed on the axis.  They are symmetric with respect
to the cloud center and move toward the center while their density increases.
They collide with each other at the center when $ J _0 $ is
close to $ J _c $.  The whole cloud becomes a thin needle when $ J _0 $
is relatively large.

     Our simulation is qualitatively similar to their simulations up to
the formation of the two condensations, one of which is the mirror image
of the other.  A condensation is produced near
the cloud edge in both simulations as far as we know from the published
figures.  Further evolution is, however, very different between their
simulations and ours.  Second generation condensations are not formed
in their simulation.  The difference comes mainly from the fact that
 the hydrodynamical equilibrium
in the radial direction can be realized only with
difficulty for an isothermal non-magnetized
gas.  Only when the mass per unit length has a certain value
$ ( = \, 2 c _s {} ^2 / G ) $ can an isothermal non-magnetized cylindrical
cloud be in equilibrium.  The difference may in part be due to the
numerical scheme.
It is difficult for an explicit code to follow further evolution after
the formation of high density cores.  Low spatial resolution may induce
the coalescence of high density condensations at an earlier stage.
We think that condensations are formed successively on the axis in
a 2D numerical simulation if the initial cloud is in a quasi-static
equilibrium and the numerical code can follow the long term evolution.

\section{APPLICATION TO TMC-1}

      In this section we briefly summarize the observations of
TMC-1 with emphasis on its star formation history.

Taurus Molecular Cloud 1 (TMC-1) is one of filamentary clouds
in the Taurus dark cloud complex,
whose distance from the Sun is about 140 pc (Elias 1978).
The filamentary structure of this cloud was first
recognized by Little and his collaborators (1978)
through mapping observations of HC$_5$N.
Since then TMC-1 has been extensively studied using
various molecular lines such as NH$_3$, CS, HC$_3$N and HC$_7$N
(T\"olle et al. 1981; Snell, Langer, \& Frerking 1982; \cite{oww88}).
It is now established that the filament is extended from SE to NW
with an apparent size of $ 17\arcmin \, \times \, 2\arcmin$,
which corresponds to a linear size of 0.68 $\times$ 0.08 pc
(\cite{hir92}; \cite{oww88}).
A significant velocity gradient (3.8 km~s$^{-1}$~pc$^{-1}$)
is observed along the minor
axis (Olano et al. 1988).  Such a velocity structure
is interpreted as possible rotation or overlapping of several filaments.
The global magnetic field in this region was studied by
Moneti et al. (1984).  They carried out optical and infrared polarimetry
in this region, and found that the magnetic field direction is from
SW to NE, being perpendicular to the filament.

This source has also been a good target for detailed studies on
chemistry in dark clouds, since a number of molecular species
are detected.  Little et al. (1979) first pointed out that
a large chemical gradient
is seen along the major axis of the filament.
They observed the NH$_3$ and HC$_3$N lines, and found that
NH$_3$ is intense in the NW part whereas HC$_3$N is intense
in the SE part.  However the origin of such a chemical gradient
has long been a puzzle.

Hirahara et al. (1992) mapped TMC-1 using the CCS
lines with a spatial resolution of $ 40 \arcsec$.
Since the optical depths of the spectral lines of CCS are
not very high, these lines are suitable for detailed studies on the
structure and chemistry of the cloud.
They reported that TMC-1 is not a
uniform filament but a chain of several dense cores spaced almost
regularly (the left panel of Figure 9).
Five cores are identified, each of which has a diameter of
0.04 - 0.10 pc.  These cores can also be seen in the
NH$_3$ map (the right panel of Figure 9).
They measured the cloud density accurately by observing
the optically thin emission of the C$^{34}$S ($J = 1 - 0$ and
$J = 2 - 1$) lines, and found that the
density tends to increase from SE to NW.
The density averaged over the
$ 40\arcsec $ beam is $4 \times 10^4$ cm$^{-3}$ at the
cyanopolyyne peak (Core D), whereas it is $4 \times 10^5$ cm$^{-3}$ at the
NH$_3$ peak (Core B).  The size and density of each core are comparable to
typical values for dense cores reported by Benson \& Myers (1989).

{}From these results, the chemical gradient seen in TMC-1 is
ascribed to a systematic chemical difference among the
dense cores.  Hirahara et al. (1992) suggested that the chemical difference
originates from the core age.  According to  gas-phase
chemical model calculations (\cite{mh90}; Suzuki et al. 1992),
carbon-chain molecules like CCS are abundant
in the early stages of cloud evolution,
while NH$_3$ is only abundant in the late stages.  On the basis of
this model, the NW cores are considered to be older than
the SE cores.  In fact, the density tends to be higher
in the NW core.  Furthermore, the {\it IRAS} source (04381+2540)
exists near the NW core, indicating that  star-formation has already
taken place.

The chemical lifetime of carbon-chain molecules is about 10$^6$ yr
according to chemical model calculations (Leung, Herbst \& Huebner
1984; Tarafdar et al. 1985; Suzuki et al. 1992).
This time scale is mainly related to the conversion time scale
from C to CO.  When a cloud is in a diffuse stage with $A_v < 2$,
the major forms of carbon are C$^+$ and C,
because molecules are mostly photodissociated
by the interstellar ultraviolet radiation.   As the cloud
becomes opaque due to gravitational contraction, photodissociation
becomes ineffective, and hence,
C$^+$ and C are consumed to form a stable molecule, CO,
through various chemical reactions.
The conversion time scale from C to CO in the opaque region is inversely
proportional to the cosmic ray ionization rate and almost independent of
the cloud density.  Because a high
abundance of C$^+$ and C is essential to the production of carbon-chain
molecules, carbon-chain molecules can only exist in the
early stages of cloud evolution.
The lifetime of carbon-chain molecules  is almost
comparable to the production time scale of NH$_3$.
Therefore, the abundance ratio between carbon-chain molecules and
NH$_3$ is a good indicator of evolutionary stages during the
10$^6$ yr after the cloud becomes opaque.

On the basis of the above scenario on chemical evolution,
the age difference between the cyanopolyyne peak position
and the NH$_3$ peak position in
TMC-1 is expected to be about 10$^6$ yr, i.e.,
the cores in TMC-1 are successively formed within
this time scale.
On the other hand, our model simulation suggests that
the propagation speed of core formation is
slightly larger than the effective sound speed.
The velocity width observed in TMC-1 is
0.4 - 0.8 km s$^{-1}$, depending on the position.  When
we adopt 0.6 km s$^{-1}$ as the effective sound speed,
the difference in ages between the cyanopolyyne peak and
NH$_3$ peak is estimated to be $ 7 \times 10 ^5$ yr.  This value is
almost consistent with the time scale estimated from the
chemical differences.  These results further support
the prediction that the cores in TMC-1 are successively formed
from NW to SE.

\section{PHYSICAL AND CHEMICAL EVOLUTION OF CORES}

      Thus far, TMC-1 is unique as a filamentary cloud indicating
sequential core formation.  Our mechanism for the sequential
core formation, however, seems to work in a filamentary molecular cloud.
The uniqueness of TMC-1 may be solely due to the fact that TMC-1
has been observed extensively in various molecular emission lines.
In this section we discuss  guidelines for searching for filamentary
clouds which may contain cores with different ages.

      As discussed in the previous section, the chemical evolution
begins as a result of the UV shielding due to fragmentation and contraction.
Let us suppose that the visual extinction is $ A _v \, \simeq \, 3 $
at the onset of the chemical evolution. This means that the penetration
of UV radiation to the core is reduced by a factor of a thousand at the
onset.  Since the visual extinction is well correlated  with
the integrated intensity of a molecular emission line, we can estimate
the brightness of a core at the onset of the chemical evolution.
Cernicharo \& Gu\'elin (1987) evaluated the correlation between
the visual extinction and
the integrated intensities
of $^{12}$CO, $^{13}$CO, and C$^{18}$O to be
\begin{equation}
I(^{12}{\rm CO}) \, = \, 5.0 \pm 0.5 \cdot (A _V - 0.5 \pm 0.2) \,
{\rm K} \, {\rm km} \, {\rm s}^{-1} \; ,
\end{equation}
\begin{equation}
I(^{13}{\rm CO}) \, = \, 1.4 \pm 0.2 \cdot (A _V - 0.7 \pm 0.3) \,
{\rm K} \, {\rm km} \, {\rm s}^{-1} \; ,
\end{equation}
and
\begin{equation}
I({\rm C}^{18}{\rm O}) \, = \, 0.28 \pm 0.5 \cdot (A _V - 1.5 \pm 0.3) \,
{\rm K} \, {\rm km} \, {\rm s}^{-1} \; ,
\end{equation}
for the Taurus region.  The core of $ A _v \, = \, 3 $ is already opaque
for $^{12}$CO, semi-transparent for $^{13}$CO and transparent for C$^{18}$O.
This means that the decay of CCS and production of NH$_3$ take place in
a dense core seen in the C$^{18}$O map.

     A core which can be identified in a
C$^{18}$O map is thought to have the a density
$ \ga \, 3 \times 10 ^4 $ cm$^{-3}$.
The dynamical time scale of such a dense core
is short,
\begin{eqnarray}
\nonumber
\tau _{dyn} \; & = & \; {1 \over \sqrt{ 2 \pi G \rho }} \; \\
& = &
\; 1.55 \times 10 ^5 \, \Bigl( { n _{\rm H _2} \over 3 \times 10 ^4 \,
{\rm cm} ^{-3} } \Bigr) ^{-1/2} \, {\rm yr} \; .
\end{eqnarray}
Core collapse and formation of a subsequent core takes
$ \sim 5 $ and $ \sim 10 $ dynamical time scales, respectively.
Since the time scale for chemical evolution is $ \sim 10 ^6 $ yr,
it is comparable to time scales for collapse and propagation of fragmentation
only when $ n _{\rm H _2} \, \ga \, 3 \times 10 ^4 \, {\rm cm} ^{-3} $.
Accordingly we can observe cores of different chemical compositions
only in a dense core which can be selected from a C$^{18}$O
map.

In fact, such a chemical difference is observed in several
clouds in the Taurus region.  Good examples are TMC-1C and TMC-1A.
These clouds are visible on the C$^{18}$O
map (Cernicharo \& Gu\'elin 1987).  TMC-1C has an elongated
structure from SE to NW with a linear size of 0.3 $\times$ 0.1 pc
(\cite{cga84}; \cite{yam93}).
The NH$_3$ map reported by Benson \& Myers (1989)
has a peak at the southeast part of the core,
whereas the CCS map has an additional peak at the northwest part
(\cite{yam93}).
Thus a significant chemical gradient can be seen along the major axis of
the cloud.  In this case, the SE core is considered to be
older than the NW core, and hence, the cores are being formed
successively from SE to NW.

A similar chemical gradient is also seen in the
TMC-1A region, as reported by Cernicharo et al. (1984).
The distribution of NH$_3$ has a peak at the infrared source position
({\it IRAS} 04365+2535).  On the other hand, the emission from
carbon chain molecules, HC$_3$N and HC$_5$N, tends to be
intense in the $ 5 \arcmin $ north position, although these molecules are
not fully mapped in this region.
Therefore, the southern part of TMC-1A seems to be in an advanced
stage of cloud evolution.
It is interesting to note that
a high-velocity outflow has recently been
detected toward the infrared source (\cite{tvm89}; \cite{hir93}),
indicating that
star formation has already taken place in the southern region.
Although a few examples of a chemical gradient in the
dense cores have been reported, as mentioned above,
more extensive observations of NH$_3$ and carbon-chain
molecules would be indispensable for observational
confirmation of the sequential core (star) formation
in filamentary clouds proposed in the present paper.

\acknowledgments

     We thank Dr. Nagayoshi Ohasi and Mr. Fumitaka Nakamura for stimulating
discussion and comments.  We also thank Dr. Lance T. Gardiner for
reading the manuscript and the anonymous referee who suggeted us to
make models 6 and 7.
This study is financially supported in part
by the Grant-in-Aid for General Scientific Research (04640270)
and Grant-in-Aid for Scientific Research on Priority Areas
(Interstellar Matter and Its Evolution 04233107).

\clearpage

\begin{table*}
\begin{center}
\begin{tabular}{llrlr}
$c _s$ & $k _{max}R$ & $ \lambda $ & $ i \omega _{max} $ & $ v _{frag} $ \\
\tableline
0.9 & 0.7559 & 8.312 & 0.5142 & 0.756 $ c _s $ \\
1.0 & 0.6436 & 9.763 & 0.4142 & 1.231 $ c _s $ \\
1.1 & 0.5345 & 11.756 & 0.3142 & 1.512 $ c _s $ \\
1.2 & 0.4225 & 14.871 & 0.2142 & 1.711 $ c _s $ \\
1.4 & 0.1008 & 62.358 & 0.0142 & 1.985 $ c _s $ \\
1.4142 & 0.0 & $ + \infty $ & 0.0 & 2.000 $ c _s $ \\
\end{tabular}
\end{center}

\caption{Dispersion Relation for 1D Model Filamentary Cloud} \label{tbl-1}
\end{table*}

\clearpage

\begin{figure}
\caption{The time evolution of model 1.  The abscissa and ordinate
are the $ z $-coordinate and time, respectively.  The density is denoted
by the darkness.  The grey scale is shown on the top of the panel.
The sound speed is taken to be $ c _s \, = \, 1.0 $.}
\end{figure}

\begin{figure}
\caption{Density and velocity distribution in model 1 at $ t \, = \, 20 $.
The upper and lower panels show the density on a logarithmic scale
and velocity in a linear scale, respectively.  The abscissa is the
$ z $-coordinate.}
\end{figure}

\begin{figure}
\caption{The same as figure 1 but for model 2 where the sound
speed is taken to be $ c _s \, = \, 1.1 $.}
\end{figure}

\begin{figure}
\caption{The same as figure 1 but for model 3 where the sound
speed is taken to be $ c _s \, = \, 1.2 $.}
\end{figure}

\begin{figure}
\caption{The same as figure 1 but for model 4 where the initial
density density increases from 0.1 to 1.0 in the region
$ -2.5 \, \le \, z \, \le 2.5 $.  The sound speed is taken to be
$ c _s \, = \, 1.0 $.}
\end{figure}

\begin{figure}
\caption{The same as figure 1 but for model 5 where the initial
density density increases from 0.1 to 1.0 in the region
$ -5 \, \le \, z \, \le 5 $.  The sound speed is taken to be
$ c _s \, = \, 1.0 $.}
\end{figure}

\begin{figure}
\caption{The same as figure 1 but for model 6 where the initial
density density increases from 0.1 to 1.0 in the region
$ -20 \, \le \, z \, \le 20 $.  The sound speed is taken to be
$ c _s \, = \, 1.0 $.}
\end{figure}

\begin{figure}
\caption{The same as figure 1 but for model 4 where the initial
density density increases from 0.1 to 1.0 in the region
$ -50 \, \le \, z \, \le 50 $.  The sound speed is taken to be
$ c _s \, = \, 1.0 $.}
\end{figure}

\begin{figure}
\caption{Total integrated intensity maps of CCS (left) and NH$_3$ (right)
emission toward TMC-1 (Hirahara et al. 1992).  The lowest contour level and
the contour interval are $ 0.4 \; {\rm K} \, {\rm km} \, {\rm s}^{-1} $.
Five cores (A - E) are identified in the filament.}
\end{figure}

\end{document}